\documentclass[apjl]{emulateapj}
\usepackage{mathptmx}

\begin{document}

\title{Dynamical and Spectral Modeling of the Ionized Gas and Nuclear Environment in NGC\,3783}
\author{Doron Chelouche\altaffilmark{1,2,3}, Hagai Netzer\altaffilmark{3}}
\altaffiltext{1} {School of Natural Sciences, Institute for Advanced Study,
		  Einstein Drive, Princeton 08540, USA; doron@ias.edu (DC)}
\altaffiltext{2} {Chandra Fellow}
\altaffiltext{3} {School of Physics and Astronomy and the Wise
                Observatory, The Raymond and Beverly Sackler Faculty of
                Exact Sciences, Tel-Aviv University, Tel-Aviv 69978,
                Israel; netzer@wise.tau.ac.il (HN)}
\shortauthors{Chelouche D. \& Netzer H.}
\shorttitle{Modeling of the X-ray outflow in NGC\,3783}

\begin{abstract}

We present a new approach for calculating the physical properties of
highly ionized X-ray flows in active galactic nuclei (AGN). Our method
relies on a detailed treatment of the structure, dynamics and spectrum
of the gas.  A quantitative comparison of our model predictions with
the 900\,ksec {\it Chandra/HETG} X-ray spectrum of NGC\,3783 shows that: (1) The highly ionized outflow is driven by thermal pressure gradients and
radiation pressure force is less important. (2) A full featured
dynamical model that provides a very good fit to the high resolution X-ray
spectrum requires a multi-phased flow with a density
power spectrum reminiscent of the interstellar medium. (3) Adiabatic
cooling is an important factor and so is an additional heating source
that  may be related to the
apparent multi-phase and turbulent nature of the flow. (4) The base of the flow is $\sim 1$\,pc from
the central object, in agreement with some, but not all previous
estimates. (5) The mass loss rate is in the range $0.01-0.1~{\rm
  M_\odot~yr^{-1}}$ which is smaller than previous estimates and  of
the same order of the mass accretion rate in this object.  

\end{abstract}

\keywords{
acceleration of particles ---
ISM: jets and outflows ---
galaxies: Seyfert ---
quasars: absorption lines ---
X-rays: individual (NGC\,3783)}

\section{introduction}

Highly ionized gas (HIG) in active galactic nuclei (AGN), also known as ``warm absorber'', was first detected in the early 1980s (Halpern
 1984) and was observationally and theoretically investigated in
 hundreds of papers (see e.g., Netzer et al. 2003, hereafter N03, and
 references therein). This component is believed to be photoionized by
 the central X-ray continuum source and is observed in some 50\% of
 all type-I (broad line) AGN (e.g., George et al. 1998, 2000, Porquet,
 Reeves, O'Brien, \& Brinkmann 2004). The poor resolution of the
 pre-{\it Chandra \& XMM} instruments did not allow full investigation
 of the HIG properties and its possible relation to the UV outflowing
 gas (e.g., Crenshaw et al. 1999). The launch of {\it Chandra \& XMM}
 allowed an in-depth analysis of the spectral properties of this gas component (e.g.,  
Kaastra et al. 2000, Kaspi et al. 2000, Branduardi-Raymont et
al. 2001, Lee et al. 2001, Steenbrugge et al. 2005). X-ray absorption lines were detected in several sources and found to be blueshifted by a few\,$\times100\,{\rm km\,s^{-1}}$
relative to the systemic velocity (though much 
higher velocities have also been reported, e.g., Chartas et
al. 2002). 

Detailed photoionization modeling of several AGN show outflows with a
stratified density and temperature structure. Such models are partly
phenomenological since they do not self-consistently account for the
flow dynamics.  Specifically, such models cannot be used to {\it
  predict} the velocity and deduce dynamically related quantities such
as the mass loss rate. This information is crucial for understanding
the effect of AGN flows on their environment (e.g., King  2003,
Scannapieco \& Oh 2004). A more complete approach requires  full,
self-consistent dynamical,  photoionization, and spectral modeling of
the flows. Today there are only a few qualitative works of this
type (e.g., Arav, Li, \& Begelman 1994, Murray et al. 1995, Chelouche
\& Netzer 2003a,b). 

This paper introduces a novel approach for investigating the
properties of highly ionized flows. It utilizes the entire spectral
information and employs state-of-the-art photoionization, dynamical,
and spectral calculations.
We apply this method to the X-ray flow in NGC\,3783 and compare our model predictions to the  900\,ksec {\it Chandra/HETG} spectrum of this source (e.g., Kaspi et al. 2002).
The paper is organized as follows:  In \S 2 we summarize the
properties of NGC\,3783 and its outflow. In \S 3 we use general
arguments to distinguish between several plausible dynamical 
models. Section 4  outlines the new formalism
and \S 5 presents the self-consistent
solution  and shows a detailed comparison with the
 X-ray spectrum of NGC\,3783. We elaborate on
the implications of our results  in \S 6 and summarize the work in \S 7.

\section{The highly ionized gas in NGC\,3783}

NGC\,3783 is a Seyfert 1 galaxy that has been studied, extensively, in
almost all  spectral bands.  It has been the focus of a
large UV and X-ray campaign with several already published papers
 (Kaspi et al. 2001, 2002, Behar et al. 2003,
Gabel et al. 2003a,b, 
Krongold et al. 2003, N03). While certainly the best studied AGN
in the X-ray band, NGC\,3783 is by no means unique and its
low resolution {\it ASCA} spectrum resembles that of other AGN
exhibiting warm absorption features (e.g., George et al. 1998). 

The bolometric luminosity of NGC\,3783, $L_{\rm tot}$, can be estimated
from its 2-10\,keV X-ray flux  reported by Kaspi et al. (2002)
and the spectral energy distribution (SED) presented in N03. This results in $L_{\rm tot}\simeq3\times10^{44}~{\rm erg\,s^{-1}}$ which we assume
to be the long-term average luminosity of the source. The mass of the black
hole, $M_{\rm BH}$, has been estimated from  reverberation mapping to
be $\sim (3\pm 1)\times 10^7~{\rm M_\odot}$ and the size of the broad
line region (BLR) $\sim 10^{16}~{\rm cm}$
(Peterson et al. 2004; see however Kaspi et al. 2000 for lower
values). Assuming solar composition for the gas we define the
Eddington ratio,
$\Gamma\equiv 1.18L_{\rm tot}/L_{\rm   Edd.}$ (e.g., Chelouche \&
Netzer 2001) where $L_{\rm Edd.}$ is the Eddington luminosity. For NGC\,3783 $\Gamma\simeq 0.1$.  

The superb 900\,ksec {\it Chandra/HETG} X-ray spectrum of NGC\,3783 allowed
the first in-depth study of its HIG outflow. The results were thoroughly
discussed in Kaspi et al. (2001, 2002), Krongold et al. (2003) and N03. Here we consider the Kaspi et al. (2002) and N03 results as a basis for our model. The main results of these papers  can be summarized
as follows: The HIG is outflowing with $v\sim 1000~{\rm km~s^{-1}}$
and the absorption line profiles exhibit extended blue wings. The velocity
(either centroid or dispersion)  does not show 
significant correlation with ionization level. The HIG consists of at
least three distinct ionization components ([low, intermediate, and
high]) whose ionization parameters, $U_{\rm ox}$ (defined as the ratio of the photon density in the energy range 0.54-10\,keV to the hydrogen number density) are ${\rm  log}(U_{ox})=[-2.4,~-1.2,~-0.6]$, 
respectively. Each component has a different column density, $N_H$,
given by, ${\rm  log}(N_H)=[21.9,~22.0,~22.3]$ for the above three
values of $U_{\rm ox}$. The N03 analysis shows
that none of the HIG components responded to the large, observed continuum variations. Recombination time
arguments suggest, therefore, a lower limit on their distance of  $[3.2,0.63,0.18]$\,pc, respectively. The line of sight covering
  factor of the flow, $C_{\rm los}$, is between 0.7 and 1.0. The
  global ($4\pi$) covering factor of the flow, $C$, obtained from the
  emission lines is $\sim 20\%$. The mass loss rate, $\dot{M}$, was estimated to be
  $\sim 75\epsilon C~{\rm  M_\odot~yr^{-1}}$, where $\epsilon$
  is the volume filling factor. For $\epsilon=1$, it  exceeds the
  expected mass  accretion rate by roughly two orders of magnitude.

\section{General model considerations}

Several models have been proposed  to qualitatively explain the acceleration of highly ionized gas in AGN (e.g., Krolik \& Begelman 1986, Balsara \& Krolik 1993, K\"{o}nigl \& Kartje 1995, Bottorff, Korista, \& Shlosman 2001, Chelouche \& Netzer 2001). The models we consider here assume that the flow is driven either by radiation pressure force, thermal pressure gradients,  or a combination of the two. The equation of motion in this case is
\begin{equation}
v\frac{dv}{dr}=\frac{1}{\rho} \left [ \frac{n_e\sigma_T L_{\rm tot}}{4\pi r^2c} \left (
  M -\frac{1}{\Gamma} \right )-\frac{dP}{dr} \right ],
\label{eqnmot}
\end{equation}
where $r$ is the distance from the ionizing source, $v$ the velocity,
 $\rho$ ($n_e$) the gas density (electron number density), $\sigma_T$ the Thomson cross-section and $P$ the thermal pressure ($P=\rho v_s^2$ where
 $v_s=\sqrt{2k_BT/\mu m_H}\simeq 166(T/10^6\,K)^{1/2}~{\rm km~s^{-1}}$
 and $\mu m_H$ is the average mass per particle; $\mu\simeq 0.6$ for fully ionized solar metallicity gas). $M$ is the force multiplier defined as the
ratio of the total radiation pressure force to that due to Compton
scattering (see Chelouche \& Netzer 2003a and references therein). For the densities associated with the
HIG ($\ll 10^{12}~{\rm cm^{-3}}$), the ionization structure of 
optically thin flows depends, to a good approximation, only on $U_{\rm
  ox}$. In this limit, $M$ is a function of $U_{\rm ox}$ and the optical depth in the line per unit thermal velocity width (e.g., Chelouche \& Netzer 2003a).

The  modeling scheme considered here combines  detailed
photoionization and spectral calculations (some aspects of which were
discussed by N03) with a simplified treatment of the gas dynamics. As shown below, the unique set of observational constraints  allows us to quantitatively investigate, for the first time, which of the aforementioned dynamical models is most relevant to the HIG outflow in NGC\,3783.

\subsection{Radiation pressure driven flows}

\begin{figure}
\plotone{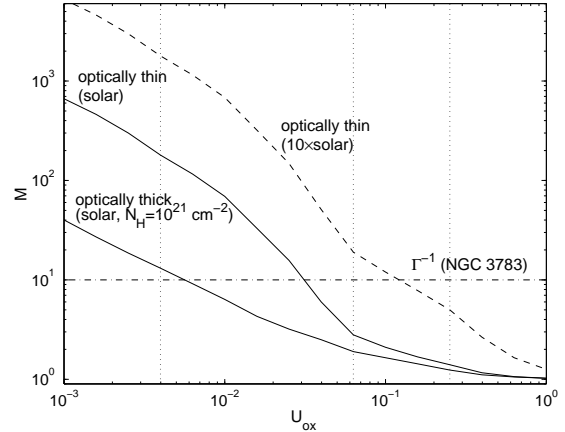}
\caption{The force multiplier, $M$, as a function of $U_{\rm ox}$ for the SED used by N03. Optically thin and thick (where $M$ is an average over the cloud) solar composition gas is shown in solid lines.  The gravitational term for NGC\,3783 is shown in horizontal dash-dotted line. The three vertical lines mark the HIG components of N03. Note that unless the flow is highly metal rich (top dashed curve), radiation pressure force is smaller than gravity for two of the three N03 solutions.}
\label{f1}
\end{figure}

Non-rotating radiation pressure dominated outflows require
$M>\Gamma^{-1}$. Figure \ref{f1} suggests that this condition is
difficult to meet for solar metallicity, optically thin  HIG, assuming
the known $\Gamma$ and SED. Specifically, this condition is not met
for the intermediate and high ionization components presented in N03,
where we find that $M<\Gamma^{-1}$ for $U_{\rm ox}> 5\times 10^{-3}$.
There is evidence that BLR metallicities can exceed solar (e.g.,
Hamann \& Ferland 1993, see however Shemmer et al. 2004). In our case,
the metallicity should exceed about $30\times$ solar for radiation pressure force to overcome gravity for $U_{\rm ox}<0.3$.  Such metalicities have never been observed in AGN. 

Radiation pressure force can exceed gravity close to the foot-point of
rotating flows. As shown by Chelouche \& Netzer (2001), HIG flows must be launched within or just outside the broad line region ($\sim 10^{16}\,~{\rm cm}$) in order to reach the observed velocities. Unless the line broadening in NGC\,3783 is unrelated to the gas dynamics, such models cannot account for the width of the line profiles given the much larger distances implied by the Behar et al. (2003) and N03 analyses. The above considerations do not take into account the effect of dust on the flow dynamics (e.g., Scoville \& Norman 1995, Everett 2002) since none has been detected in NGC\,3783. We  conclude that radiation pressure force is unlikely to drive the flow in NGC\,3783 to the observed velocities.

We note that HIG flows may not be steady-state phenomena. In this case, the HIG may have been accelerated in the past (when $M$ was larger or the object brighter) and we observe it now at its coasting or even decelerating phase. Although plausible (e.g., Chelouche \& Netzer 2001), such a model adds little to our understanding of the general phenomenon.

\subsection{Thermal pressure driven flows}

Acceleration by thermal pressure gradients (see Parker 1958) has been suggested to drive extremely ionized high temperature gas in AGN (e.g., Begelman, McKee, \& Shields 1983, Krolik \& Begelman  1986, Balsara \& Krolik 1993 and Woods et al. 1996). In particular, Balsara \& Krolik (1993) and Woods et al. (1996) presented the results of detailed, time-dependent hydrodynamic calculations of thermal pressure driven flows for a wide range of initial conditions. The conclusion is that highly ionized gas near its Compton temperature may be driven to terminal velocities of up to $\sim 2000~{\rm km~s^{-1}}$. Here we investigate the possibility that such a mechanism  drives the ionized gas in NGC\,3783.

For a spherically expanding flow,
\begin{equation} 
\rho \propto r^{-2}v^{-1}.
\label{cont}
\end{equation}
In the limit of $M\ll\Gamma^{-1}$ (see figure \ref{f1}), and assuming a polytropic flow with index $\gamma$ ($\gamma=1$ for isothermal gas) we can rewrite equation \ref{eqnmot} as
\begin{equation}
\frac{1}{v}\frac{dv}{dr}=\frac{1}{v^2-\gamma v_s^2}\left [
  \frac{2\gamma v_s^2}{r}-\left ( \frac{n_e\sigma_TL_{\rm tot}}{4\pi \rho
      r^2c\Gamma} \right )
  \right ].
\label{eqnmot1}
\end{equation}
Super-sonic steady-state winds that are launched sub-sonically should cross the sonic point at $r_c={GM_{\rm BH}}/{2\gamma v_s^2}$, where $v(r_c)=\sqrt{\gamma} v_s(r)$. At this point both the numerator and the denominator vanish thus the equation of motion is defined at all radii. Noting that $M_{\rm BH} \propto L_{\rm Edd.}\propto L_{\rm tot}/\Gamma$ we find 
\begin{equation}
r_c\simeq \gamma^{-1}\left (\frac{L_{\rm tot}}{10^{44}~{\rm erg~s^{-1}}}\right ) \left (
  \frac{10^6\,{\rm K}}{T} \right ) \left ( \frac{0.1}{\Gamma} \right )~{\rm pc}.
\label{evap}
\end{equation}

\begin{figure}
\plotone{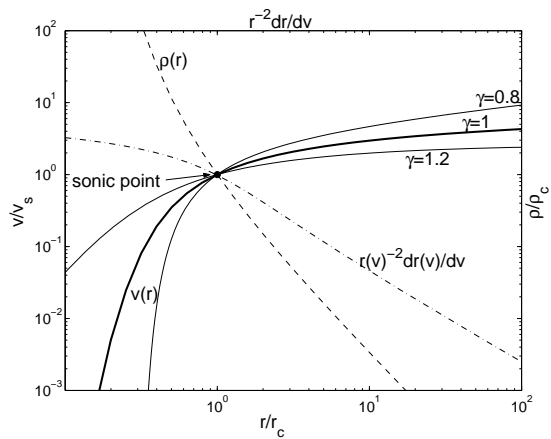}
\caption{Normalized velocity profiles of thermal pressure driven
polytropic  flows. $r_c$ is the location of the sonic point (see
text). Larger $\gamma$ (i.e., cases where adiabatic cooling is more
important) result in lower velocity for the same sonic point. Also
shown is the density profile for a $\gamma=1$ isothermal flow (dashed
line) which resembles the hydrostatic density profile for $r<r_c$. The
column density per unit velocity for an isothermal flow at a fixed
ionization parameter ($\propto r(v)^{-2} dr/dv$; dash-dotted curve) illustrates the way the opacity decreases with velocity (see \S 4.3)}
\label{f6}
\end{figure}

For a given $\gamma$, wind solutions are self-similar and scale with $v_s$ and $r_c$. Several such solutions for different values of $\gamma$ are shown in figure \ref{f6}. The terminal velocity of such flows is of the order of (or a few times larger than) the sound-speed at the critical point (i.e., the escape velocity). For isothermal flows  ($\gamma=1$) the terminal velocity is a few times the sound-speed ($\sim 4v_s$ at $r\sim 10^2r_c$) while for polytropic flows with $1<\gamma<5/3$ the terminal velocity is lower due to adiabatic losses (e.g., Lamers \& Cassinelli 1999). Polytropic flows where $\gamma<1$ imply excessive heating and result in higher flow velocities.

Also shown in figure \ref{f6} is the density profile for an isothermal flow. For $r<r_c$ the density profile is similar to the hydrostatic case. The density beyond the sonic point declines more rapidly compared to the hydrostatic case. The column density of the flow, from some $r$ to infinity, is
\begin{equation}
N_H(r)=\int_r^\infty dr'n_H(r').
\label{colu}
\end{equation}
For super-sonic isothermal flows,  $v\propto \sqrt{{\rm ln}(r)}$ and  $N_H\simeq rn_H(r)$. 

Returning to NGC\,3783, we notice that the temperature of the high
ionization component found by N03, combined with the mass of the
central object in this source and assuming $\gamma=1$ gives (equation
\ref{evap})  $r_c\sim 3$\,pc. This distance is consistent
with the three lower limits on the distance found by N03. Using equation
\ref{colu} and the density of the hot component found by N03, we get
$N_H \sim 10^{22}~{\rm cm^{-2}}$ which 
is similar to the column density of this component obtained by
N03. Thus, the conditions in this source are consistent with a picture
where the highest ionization gas is in a state of thermal pressure
driven flow. As shown below, there are other observational constraints
that are satisfied by such a model.

\subsection{The density structure of the flow}

The continuity condition (equation \ref{cont}) implies $U_{\rm ox} \propto v$. This has been shown to result in high ionization absorption line profiles that are more blueshifted and broader compared to low ionization lines (Chelouche \& Netzer 2002).  Such trends are not observed in NGC\,3783 (Kaspi et al. 2002). In fact, lines of different ionization states are remarkably similar which indicates similar dynamics. It is unlikely that different ionization components are launched with different initial conditions and end up showing the same line profiles. A more likely explanation is that all flow components  are co-spatial and are accelerated together to their observed velocities (a similar conclusion has been reached by N03 based on the apparent pressure equilibrium between the different phases).

The co-existence of several gas phases is a well known phenomenon in
the interstellar medium (ISM) and in molecular clouds. A multi-phase
gas naturally occurs for a turbulent medium where the relation between
the density $\rho$ and the physical scale $\xi$ follows a powerlaw
distribution, $\rho \propto \xi^\beta$ (see \S 4 for the full
definition).  Such dependences have been observed on large ($1-10^3$\,pc; e.g., Elmegreen \& Scalo 2004) as well as small ($10^{15}-10^{18}$\,cm; Deshpande 2000) scales. 

Consider a small section of the flow over which both $r$ and $v$ are constant. Since $N_H(\xi)\propto \rho \xi \propto \xi^{\beta+1}$ and $U_{\rm ox}\propto \rho^{-1}$, we get $N_H\propto U_{\rm ox}^\alpha$ where $\alpha=-(\beta+1)/\beta$ (cf. Krolik \& Kriss 2001). Thus, there is a simple local relationship between the column density and the ionization parameter involving the size distribution parameter of the medium. This dependence holds also for geometrically thick flows and the integrated column density provided the self-similar scaling holds for all $r$ (see \S 4.3). 

Returning to NGC\,3783, we find that the N03 three component solution
implies $\beta\sim -0.8$. Such a value of $\beta$ usually
characterizes turbulent media like the ISM and
molecular clouds and is consistent with recent simulations (e.g., Boldyrev, Nordlund, \& Padoan 2002).

Our model assumes the following range of densities at every location $r$ 
\begin{equation}
\rho=\rho_{\rm min}\left ( \frac{\xi}{\xi_{\rm max}} \right )^\beta,
\label{rho}
\end{equation}
where $\rho_{\rm min}$ and $\xi_{\rm max}$ are the (location dependent)
minimum density and maximum length, respectively, and $\beta$ is a constant to be determined later. We require that the surface of a sphere with radius $r$ is fully covered by all phases in the density range $[\rho_{\rm min},\rho_{\rm max}]$, 
\begin{equation}
\int d\xi(\rho)^2=4\pi r^2.
\label{covxi}
\end{equation}
This results in
\begin{equation}
\xi_{\rm max}=2\sqrt{\pi}r\left [1- \left (\frac{\rho_{\rm
        max}}{\rho_{\rm min}} \right
  )^{2/\beta}  \right ]^{-1/2}
\label{normxi}
\end{equation}
where $\xi_{\rm max}$ is defined over the surface of the sphere. Assuming that the density contrast, $\rho_{\rm max}/\rho_{\rm min}$, is a constant of motion then $\xi_{\rm max}\propto r$.

\begin{figure}
\plotone{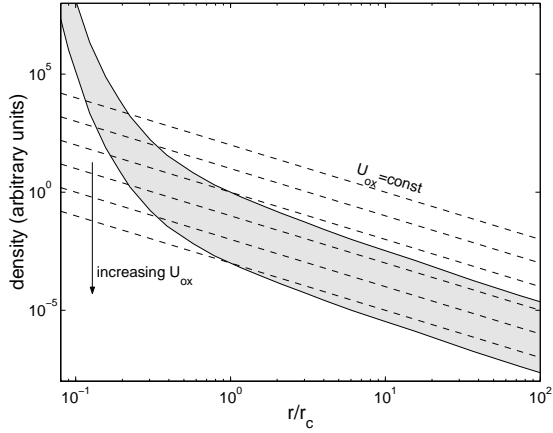}
\caption{The density profile of a multi-phase flow with a fixed
  density contrast. Also shown are lines of constant ionization
  parameter. Clearly, the expansion of an optically thin (to continuum absorption) flow results in an increase of its mean level of ionization.}
\label{f11}
\end{figure}

The mass loss rate due to all flow components can be calculated by
defining the mean density over a (thin) spherical shell,
\begin{equation}
\left < \rho (r) \right > = \frac{1}{4\pi r^2} \int
d\xi^2 \rho(\xi;r),
\label{rhoav}
\end{equation}
which may be expressed, using the above relations, as
\begin{equation}
\left < \rho (r) \right >=\frac{\rho_{\rm min}}{2\pi (\beta+2)} \left (
  \frac{\xi_{\rm max}}{r} \right )^2 \left [ 1-\left ( \frac{\rho_{\rm
        max}}{\rho_{\rm min}}
  \right )^{(\beta+2)/\beta} \right ].
\label{rhoav1}
\end{equation}
The mass loss rate is then given by
\begin{equation}
\dot{M}= 4\pi r^2\left < \rho(r) \right > v =\frac{8\pi r^2\rho_{\rm min} v}{\beta+2}\frac{1-(\rho_{\rm max}/\rho_{\rm min})^{(\beta+2)/\beta}}{1-(\rho_{\rm max}/\rho_{\rm min})^{2/\beta}}.
\label{dotm1}
\end{equation}
This conforms with the mass loss rate of a single phase spherical flow in the limits $\beta\rightarrow 0,~\rho_{\rm max}= \rho_{\rm min}$. Assuming a divergence free flow,  mass conservation implies
\begin{equation}
\rho_{\rm min}r^2v={\rm const.}
\label{cont2}
\end{equation}
This continuity condition is identical in form to that of a single-phase
medium. It means that the highest ionization parameter, $U_{\rm ox}^{\rm max}$, corresponding to the lowest density $\rho_{\rm min}$, is proportional to the velocity. By construction there are fixed density and ionization parameter
contrasts at every location ($\rho_{\rm min}/\rho_{\rm max}=U_{\rm ox}^{\rm min}/U_{\rm ox}^{\rm max}$). An example of a 
density profile of a multi-phase flow is shown in figure \ref{f11} together with several iso-ionization parameter curves. Clearly, the mean level of ionization increases as the gas expands. 

\begin{figure}
\plotone{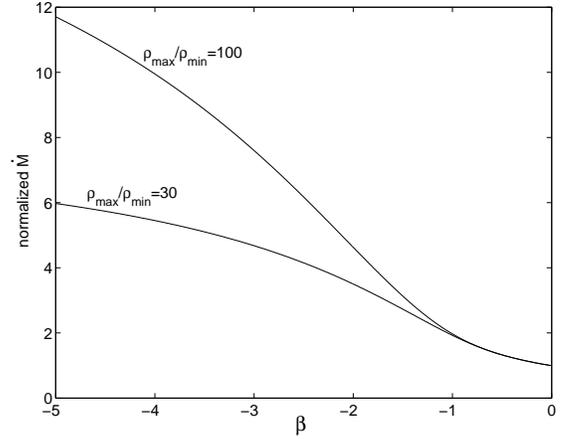}
\caption{The mass loss rate, $\dot{M}$, as a function of the density powerlaw index $\beta$ for different values of the density contrast (equation \ref{dotm1}). $\dot{M}$ increases with decreasing $\beta$ and increasing density contrast since the fraction of the denser regions of the flow is larger in those cases (see text).}
\label{mdot}
\end{figure}

For a given velocity profile, the mass loss rate depends on the density contrast and $\beta$. For a fixed $\rho_{\rm min}$, smaller $\beta$ and/or larger density contrasts result in denser regions having larger scales and a larger contribution to $\dot{M}$. The differences are small for $\beta >-1$ since the mass fraction of the denser regions is negligible (figure \ref{mdot}).

The effect of inhomogeneity on the flow dynamics can be approximated
by noting that all components in the N03 solution are in a rough pressure equilibrium. Under such conditions one must solve for the combined dynamics of all phases since their coupling is strong and the momentum is quickly (relative to the dynamical timescale) distributed throughout the flow. The dynamical effect can therefore be included by considering their contribution to the inertial mass. The mean (over volume) density  is $\sim 3/(\beta+3)\rho_{\rm min}$. In this approximation, the critical point is given by 
\begin{equation}
r_c=\left [ (1+\beta/3)\gamma \right ]^{-1}\frac{GM_{\rm BH}}{2 v_s^2}, ~~~
v(r_c)=v_s \sqrt{(1+\beta/3)\gamma}.
\end{equation}

In the above analysis we have neglected the effect of drag forces between the different phases. This is justified if such condensations are constantly forming and evaporating on short timescales compared to the dynamical time and the mean net effect is averaged-out. Thus, the flow considered here  is very different from other works in which discrete, dynamically independent clouds, rather than time-dependent condensations, are assumed (e.g., Chelouche \& Netzer 2001, Everett 2002). 

\subsection{Thermal balance in a multi-phase expanding medium}

As demonstrated in several works (e.g., Begelman, McKee, \& Shields
1983, Balsara \& Krolik 1993, Woods et al. 1997), adiabatic cooling
can have an important effect on the thermal equilibrium and, hence, on
the dynamics of continuously expanding Compton-heated winds in low
luminosity AGN. Here we investigate the effect of such cooling on the
thermal structure of the multi-phased HIG in NGC\,3783  where the  temperature is below the Compton temperature and the medium may be turbulent. 

\begin{figure}
\plotone{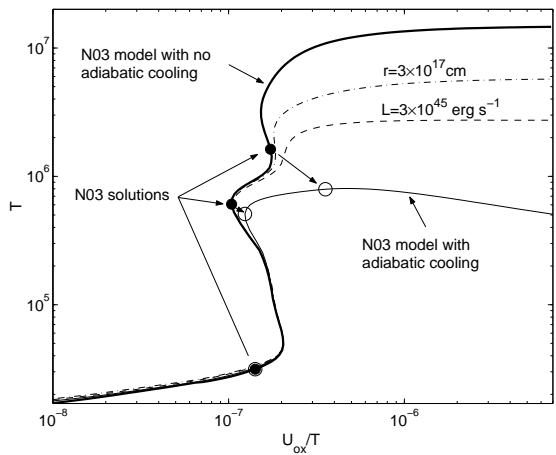}
\caption{The effect of adiabatic cooling on the stability curve of
  photoionized gas in NGC\,3783. In all
  models the gas is assumed to flow with a velocity of $1000~{\rm
    km~s^{-1}}$, the distance is $10^{19}$\,cm,  and the SED and
  luminosity ($3\times10^{44}~{\rm erg~s^{-1}}$) are those defined in
  N03 (see \S 2). The thick solid line is 
  the N03 stability curve with their assumed, $r=10^{19}~{\rm cm}$ and
  no adiabatic cooling. The thin solid line has the same
  parameters but includes adiabatic cooling. Additional
  stability curves that include adiabatic cooling were calculated by varying  one parameter at a time: $r,~L$. The
  parameter that was changed is denoted next to the curves.}
\label{cool}
\end{figure}

Neglecting the small contribution due to velocity gradients, the adiabatic cooling term for a spherically expanding continuous flow is
\begin{equation}
\Lambda_{\rm adiabatic}\simeq 2\rho v_s^2\frac{v}{r}.
\label{adicool}
\end{equation}
The N03 model of NGC\,3783 did not include adiabatic
cooling. Considering the observed properties of NGC\,3783 
($r=10^{19}~{\rm cm}$ and  $v\simeq 1000~{\rm km~s^{-1}}$) and
including the process in the ION thermal calculations, we find  that
adiabatic cooling lowers the maximum temperature by 
more than an order of magnitude (see figure \ref{cool}). Since we do
not have a numerical hydrodynamic code, we have tested this result
against the numerical calculations of Woods et al. (1996) under
similar conditions (such as the Eddington ratio, SED, and gas
velocity; see their figure 14) and find them to be in good
agreement. Thus, the numerical approach 
adopted here gives the correct temperature and expansion velocity
which is required by our calculations. The effect is smaller when the
ionizing flux is higher since heating is proportional to $L_{\rm
  tot}/r^2$ (see the individual curves in figure \ref{cool}). The ionization structure of the flow and, hence $M$, are less affected by adiabatic cooling. 

Adiabatic cooling also alters the  stability curve for the gas. The
three N03 solutions still lie on the stable branches of the stability curve yet the pressure of the high ionization component is smaller by roughly a factor 3 so that strict pressure equilibrium between all components is not maintained.

The flow considered here is multi-phased and  turbulent (see \S
6.3). In such a case there is an additional internal energy source
entrained in the large eddies. This energy is transferred to smaller
and smaller scales until it dissipates (e.g., via Coulomb collisions)
and serves as an additional heat source for the gas. The corresponding
heating rate is
\begin{equation}
G_{\rm turbulence}\simeq \rho \frac{v_{\rm turb}^3}{r}
\label{turbheat}
\end{equation}
where $v_{\rm turb}$ is the turbulence velocity (see Bottorff \&
Ferland 2002). Typically, $v_{\rm turb}$ is of the order of $v_s$ and
the flow velocity, $v$ (e.g., Shu 1992). The size of the largest eddy is comparable to the flow extent and, hence, to $r$. Hence, heating by turbulent dissipation can offset the adiabatic cooling. Below we
consider several models where the two terms play different roles.

\section{A Thermal Pressure Driven Flow Model for NGC\,3783}

In \S 3.2 we have shown a qualitative agreement between the
predictions of a multi-phase, thermal pressure driven flow model and
the observations of NGC\,3783. Section 3.3 outlined the physical model in detail. Here we carry out a more quantitative investigation of the model and 
apply it to the outflowing gas in NGC\,3783. The main aim is to give 
detailed spectral predictions and compare them to the high resolution X-ray
observations. We use the simplified, steady-state treatment of the
flow dynamics presented in \S 3.2  and  a sophisticated scheme for the
calculation of the transmitted spectrum through the flow. As
explained, the changes in the gas temperature and the attained
velocities are in general agreement with more sophisticated hydro-calculations.

Here the flow is assumed to evaporate from some mass reservoir (e.g., the accretion disk or the putative torus), expands as it is accelerated to cover a fair fraction of the AGN sky, and crosses our line-of-sight at some finite radius (to be determined later). Such  flow configurations are often assumed in 1D modeling of AGN flows (e.g., Murray et al. 1995) and have been qualitatively confirmed by detailed hydrodynamical simulations of energy driven flows (e.g., Balsara \& Krolik 1993; see also Woods et al. 1996).

\subsection{Calculation Scheme}

We begin the calculation by specifying the initial 
 conditions. These include the location, $r_0$, where
 the flow enters our line-of-sight (and need not be equal to
$r_c$),  $U_{\rm ox}^{\rm max}(r_0)$, and $U_{\rm ox}^{\rm min}(r_0)$ which also determine the density contrast throughout the flow. An additional parameter is the density powerlaw index, $\beta$. The other parameters, $L_{\rm   tot}$  and $\Gamma$, are fixed at their observed values.

The next step involves photoionization and thermal calculations of non-LTE gas in statistical equilibrium (this is justified since the dynamical timescales are larger than the recombination timescales by several orders of magnitude; Krolik \& Kriss 2001, N03). These calculations are carried out using ION\,2004, the 2004 version of the ION photoionization code (e.g., N03, with the addition of including new di-electronic recombination rates for iron; see Netzer 2004) which includes all important heating and cooling mechanisms and is suitable for exploring the physics of HIG in photoionization equilibrium. The code includes also the effect of adiabatic cooling on the thermal structure of the highest ionization phase (see equation \ref{adicool}). Once the temperature profile of the high ionization phase is calculated, we approximate it by a polytropic relation ($T\propto \rho^{\gamma-1}$) which provides a good approximation for the cases considered here, and solve the equation of motion. (We note that, for the relevant parameter space,  radiation pressure force acting on the low ionization phases of the gas has a negligible effect on the global flow dynamics). The process iterates several times  until $v(r)$ and $\gamma$ converge.

The next step is to calculate the ionization and thermal structure of all flow phases at location $r$. This is done for a flow section of length $\delta r$ in the radial direction. $\delta r$ is taken as the scale over
which the flow velocity changes by $10\,{\rm km~s^{-1}}$. This ensures
high resolution spectral predictions for the absorption lines. For this
purpose we use a discrete version of the continuous density distribution
(equation \ref{rho}) that includes 50 phases corresponding to roughly 0.1\,dex intervals in $U_{\rm  ox}$.  When calculating the
transmitted spectrum through sections of the flow we assume that
equation \ref{normxi} holds also for the line-of-sight direction albeit with a
different normalization (spherical condensations or ``clouds'' are not
the general case under such conditions; e.g., Blumenthal \& Mathews
1979). This means that
\begin{equation} 
\rho = \rho_{\rm min} (\xi^l/\xi^l_{\rm max})^\beta
\label{xil}
\end{equation}
where $\xi^l$ is the length scale in the radial direction and 
\begin{equation}
\xi^l_{\rm max}=\delta r \left [ 1- \left ( \frac{\rho_{\rm
        max}}{\rho_{\rm min}} \right )^{1/\beta} \right ]^{-1}.
\label{xil2}
\end{equation}
$\xi^l$ is determined by the somewhat arbitrarily defined $\delta r$ and is therefore independent of $\xi$.   The ratio $\xi/\xi^l$ is constant, by construction, for all phases, and is their geometrical aspect ratio (e.g., Blumenthal \& Mathews 1979). This arbitrariness in the shape and the number of condensations (see the appendix) has no bearing on the results presented in this work which rely on observationally deduced quantities, and is related to our ignorance concerning the hydrodynamical properties of such flows and the exact shape (e.g., the number of absorption systems) of the X-ray line profiles.

The contribution of all phases to the line and continuum opacity at
each section of the flow is calculated assuming full line-of-sight
coverage of the continuum source (see below). The transmitted continuum then serves  
as the ionizing spectrum for the photoionization calculation of the successive
section at $r+\delta r$. The velocity shifts at each point are taken
into account when calculating the transmitted spectrum. The
calculation continues to large $r$ and produces a high resolution,
theoretical X-ray spectrum. 

The global covering fraction of a phase with density $\rho$
in the range $[\rho_{\rm min}(r),\rho_{max}(r)]$  at
a distance $r$ is given by
\begin{equation}
C(\rho,r)\simeq \frac{\xi(\rho)^2}{4\pi r^2} \frac{\delta
  r}{\xi^l(\rho)}\sim \left ( \frac{\rho}{\rho_{\rm min}(r)} \right
)^{1/\beta}.
\label{cg}
\end{equation}
Clearly, the covering factor, expressed in this way, can be much
smaller than unity for the denser phases and can significantly effect
the transmitted 
continuum through the flow. However, this is only a geometrical
definition and the more relevant quantity is the velocity dependent covering
factor, $C(\rho,v)$,  which
takes into account the gas motion and thus 
enables the proper treatment of the opacity (e.g., Gabel et
al. 2003).  We note that, in a turbulent flow, there is no
one-to-one correspondence between $v$ and $r$. Turbulence broadens the
lines and increases $C(\rho,v)$ (see also \S 5).

\subsection{Dependence on initial conditions}

\begin{figure}
\plotone{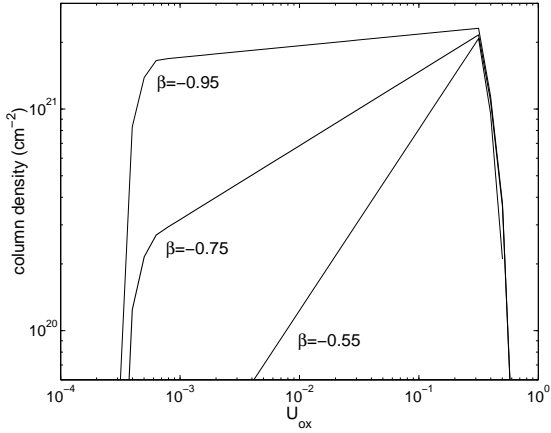}
\caption{The effect of $\beta$ on the total column density distribution as a function of ionization level (i.e., $U_{\rm ox}$). The linear part of the curve is sensitive to $\beta$ (see text). Steeper density profiles (smaller $\beta$) result in the flow being dominated by low ionization gas.}
\label{f4a}
\end{figure}

The solution of the self-consistent model depends on several
initial and boundary conditions. The
dependence on the initial conditions of the
problem, $\beta, U_{\rm ox}^{\rm max},~U_{\rm ox}^{\rm
  min}$, and $r_0$ is discussed here for the case of isothermal flows. Qualitatively similar results are obtained also for polytropic flows. 

\begin{figure}
\plotone{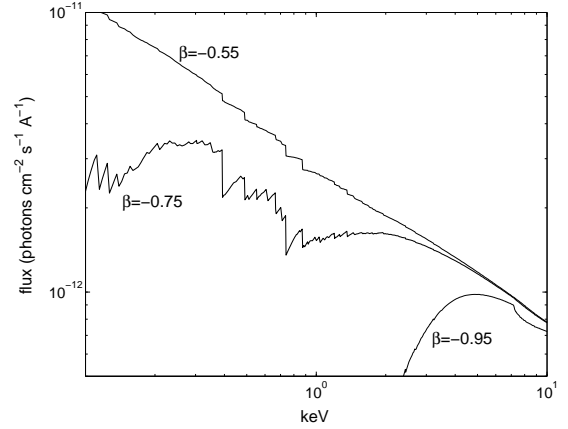}
\caption{The effect of $\beta$ on  the transmitted spectrum. Steeper density profiles (smaller $\beta$) result in low ionization gas  contributing more to the gas opacity with a stronger 
  suppression of the soft X-ray flux.}
\label{f4b}
\end{figure}

We first  explore the dependence of the column density distribution on $U_{\rm ox}$ and $\beta$. As discussed in \S 3.3 and shown in
figure \ref{f4a}, $\beta$ affects the column density distribution such that smaller $\beta$ results in larger columns of low ionization gas. The expansion of the flow and the imposed boundary conditions result in rapid decline of the column density for extreme values of $U_{\rm ox}$. Different column densities result in different transmitted spectra.
This is shown in figure \ref{f4b} where we find that smaller $\beta$ result in more opaque flows. The effect can be dramatic even for 0.2 change in $\beta$ in cases of large integrated columns. 

Smaller density contrasts, i.e., larger $U_{\rm ox}^{\rm min}$
at a fixed $U_{\rm ox}^{\rm max}$,  result in smaller columns of less
ionized gas (figure \ref{f4a}) and therefore lower opacity  and
higher transmitted soft X-ray flux (i.e., softer spectra, figure \ref{f4b}). 

The distance, $r_0$, where the flow crosses our line-of-sight with velocity $v_0$, affects the total column density of the flow. The effect is manifested as a change in the X-ray ``color''. For example, smaller $r_0$ imply larger densities, more opaque gas, and harder X-ray spectrum. This is especially important if our line-of-sight crosses the subsonic section of the flow  ($r_0<r_c$) given the nearly hydrostatic structure of the gas.

\begin{figure}
\plotone{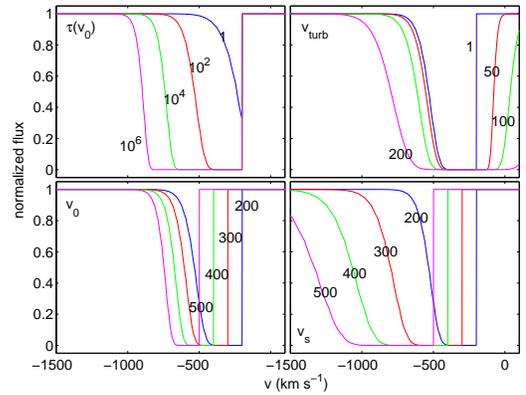}
\caption{Absorption line profiles as a function of $\tau(v_0)$ (upper-left), $v_{\rm turb}$ (upper-right), $v_0$ (lower left), and $v_s$ (lower-right; here, $v_0=v_s$ is assumed). The model marked by a blue line in all panels corresponds to $\tau(v_0)=10^2,~v_{\rm turb}=1,~v_0=200\,{\rm km~s^{-1}}$, and $v_s=200\,{\rm km~s^{-1}}$. All line profiles exhibit an extended blue wing.}
\label{f5}
\end{figure}

We next consider the effect of initial conditions on the
absorption line profiles and assume, for demonstrative purposes, that lines are produced throughout the flow by a single ionization phase (i.e., over a narrow range in $U_{\rm ox}$). In such a case, the optical depth as a function of velocity (for a phase with a given $U_{\rm ox}$) is 
\begin{equation}
\tau(v)\propto \int_0^\infty dv'r(v')^{-2} \left (\frac{dr(v')}{dv'} \right ){\rm  exp} \left ( \frac{v-v'}{2v_{\rm turb}^2} \right )
\label{tauv}
\end{equation}
where we have assumed a Gaussian kernel for
line broadening due to e.g., a turbulent velocity field. The dependence of the line profile on $\tau(v_0)$ (the optical depth at $v_0$), $v_{\rm turb}$, and $v_s$ is shown in figure \ref{f5}. The calculations show that all line profiles
exhibit extended blue wings due to the flow expansion with
the most extreme line asymmetries occurring for optically thin lines
(e.g., $\tau(v_0)\sim 1$). Larger $v_{\rm turb}$  broadens the lines
and makes them more symmetrical with respect to $v_0$. $v_0$ naturally
causes a line shift such that larger $v_0$ will result in more
blueshifted, narrower lines for the same $v(r)$. When calculating the
effect of the sound-speed (of the highest ionization component),
$v_s$, on the line profile we have assumed $v_0=v_s$ which is  
 relevant to super-sonic flows that are optically thin in the continuum. As shown, larger $v_s$ imply larger
terminal velocities and considerably broader lines. Obviously, the observed line profiles may have a more symmetrical shape depending on the spectral resolution of the instrument.

\section{Fitting the 900\,ksec Chandra spectrum of NGC\,3783}

We resort to an iterative scheme to fit the observed high resolution
X-ray spectrum of NGC\,3783. We first try to constrain the flow
dynamics. This is accomplished  by taking the theoretical line profiles, 
normalizing the column densities to agree with those obtained by N03,
and convolving the calculated profile with the known instrumental
profile (Kaspi et al. 2002). A comparison between the predicted and
the observed line profiles defines a narrow range of $r_c$ (hence,
$v_s$ and conversely $U_{\rm ox}^{\rm max}$), $v_0$ and $v_{\rm
  turb}$. An initial estimate for $\beta$ 
comes from the scaling of the column density with ionization parameter
(see \S 3.3).   This allows us to employ the next step in our fitting
scheme which is a more systematic and time consuming search in a
localized region of the parameter space, involving complex spectral
calculations (section 4.2). 

The calculated line and continuum spectra
take into account the probability for 
our line-of-sight to intersect any given phase at every location in the flow 
(see \S 4.2). We assume that the finite thermal and turbulent
broadening results in $C(\rho,v)=1$ (the opacity at every velocity bin
is properly scaled so that the integrated column density is
conserved). The best-fit model is obtained by 
comparing the detailed theoretical spectrum to the observed spectral
shape and to individual absorption line profiles. All calculations
assume a photon continuum slope, $\Gamma=1.6$, which
best characterizes the hard X-ray spectrum of this source.

\begin{figure}
\plotone{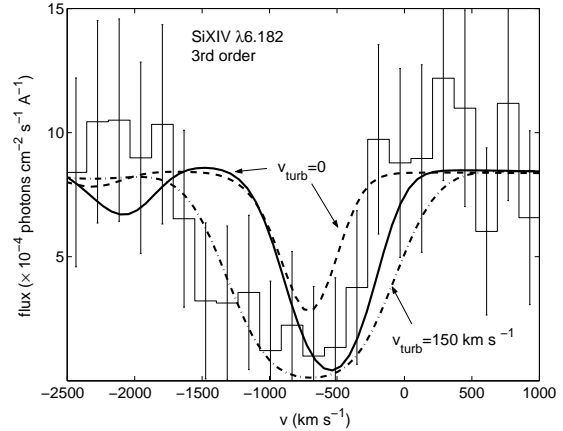}
\caption{Comparison of the calculated \ion{Si}{14} $\lambda
  6.182$ line profile with the third order MEG data of NGC\,3783. The
  solid line shows an adiabatic model which crosses our line-of-sight
  at its sonic point at $0.5$\,pc. Such a model cannot account for the
  observed blueshift or width of the line. A non-turbulent flow that
  crosses our 
  line-of-sight at $1.5$\,pc has a more blueshifted centroid but a
  small equivalent width (dashed line). A similar but turbulent
  ($v_{\rm turb=150~{\rm km~s^{-1}}}$) flow provides a better fit to
  the data (dash-dotted line).} 
\label{f22}
\end{figure}

Considering only radiative heating and cooling, and adiabatic cooling,
we find a set of models (hereafter adiabatic models) which are
consistent with the spectral shape and whose critical point is about
$0.5$\,pc from the center and $r_0=r_c$. The density contrast of the
flows spans more than four orders of magnitude at every location
(e.g., $U_{\rm ox}^{\rm min}\sim 10^{-4},~U_{\rm ox}^{\rm max} \simeq
0.4$ at $r_0\sim 1$\,pc). Such flows reach a velocity of about
$600~{\rm km~s^{-1}}$ and the absorption line profiles, convolved with
the instrumental resolution, are shown in figure \ref{f22}.  Clearly,
the predicted lines are narrower and less blueshifted than observed. A better fit is obtained by requiring that the flow crosses our line of sight at large distance ($\sim 1.5$\,pc) and the line profiles are broadened by a turbulent
 velocity field. For \ion{Si}{14}\ $\lambda 6.182$, a $v_{\rm turb}\sim 150~{\rm km~s^{-1}}$ provides a better fit to the data (figure \ref{f22}). 

The observational indication for the existence of turbulence suggests that the thermal equilibrium of the gas may be affected by it. Due to the uncertainties associated with the combined effects of adiabatic cooling and turbulent dissipation heating we study a case where $\Lambda_{\rm adiabatic}=G_{\rm turbulence}$ and radiative processes are the only important heating and cooling mechanisms. In this case, the high ionization, volume filling  phase of the flow may be considered isothermal (e.g., Begelman, McKee, \& Shields 1983). We refer to it as the isothermal model but note that all components which are not fully ionized have radial temperature dependence.

\begin{figure*}
\centerline{\includegraphics[width=14cm]{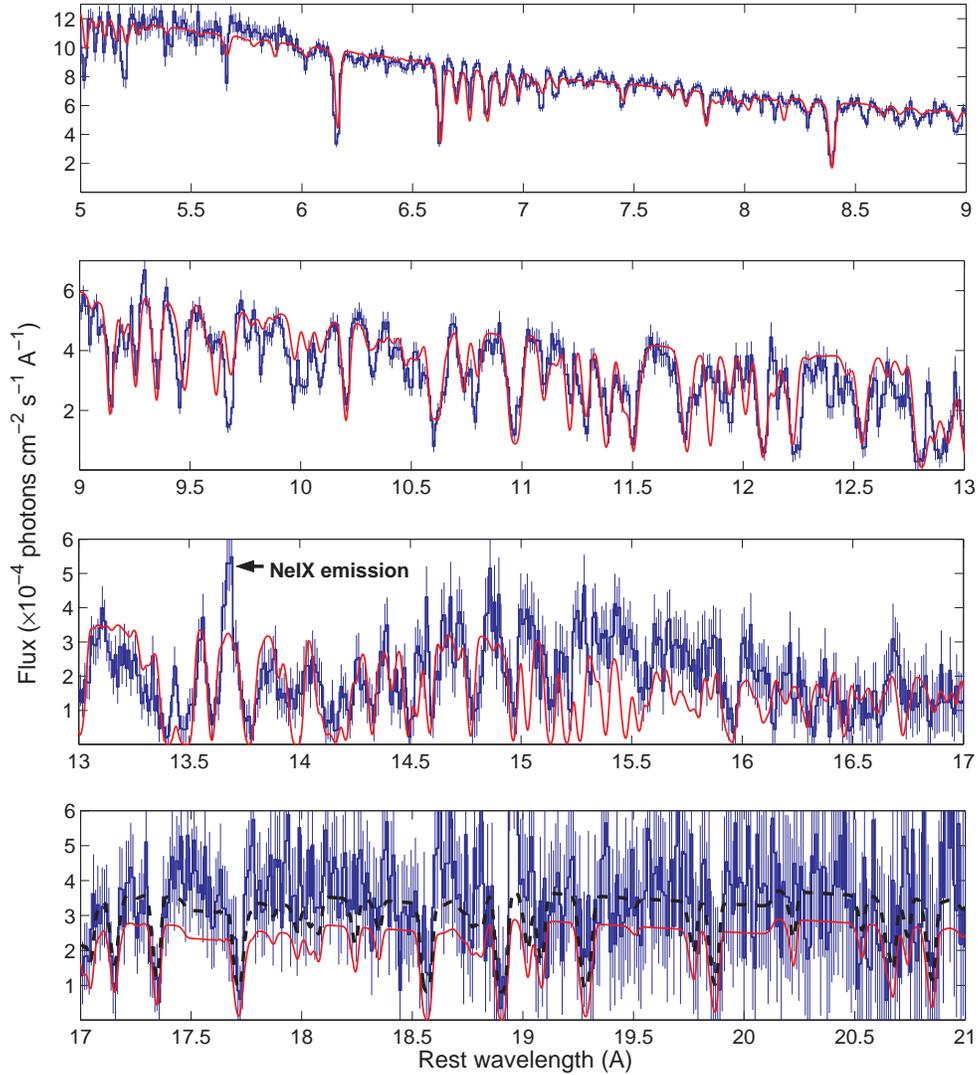}}
\caption{Full fit to the 900\,ksec {\it Chandra/HETG}  X-ray
  spectrum of NGC\,3783. This is a pure absorption model and
   line and continuum emission is not included. The model
  provides an acceptable fit over most bands but does not reproduce well the shape of the UTA near 15-16\AA. The solid line assumes full line-of-sight coverage of the source and the dashed line in the bottom panel an 80\% coverage.  The mismatch near 5.5\AA\ is due to a well known calibration effect (N03).}
\label{f14}
\end{figure*}

\begin{figure}
\plotone{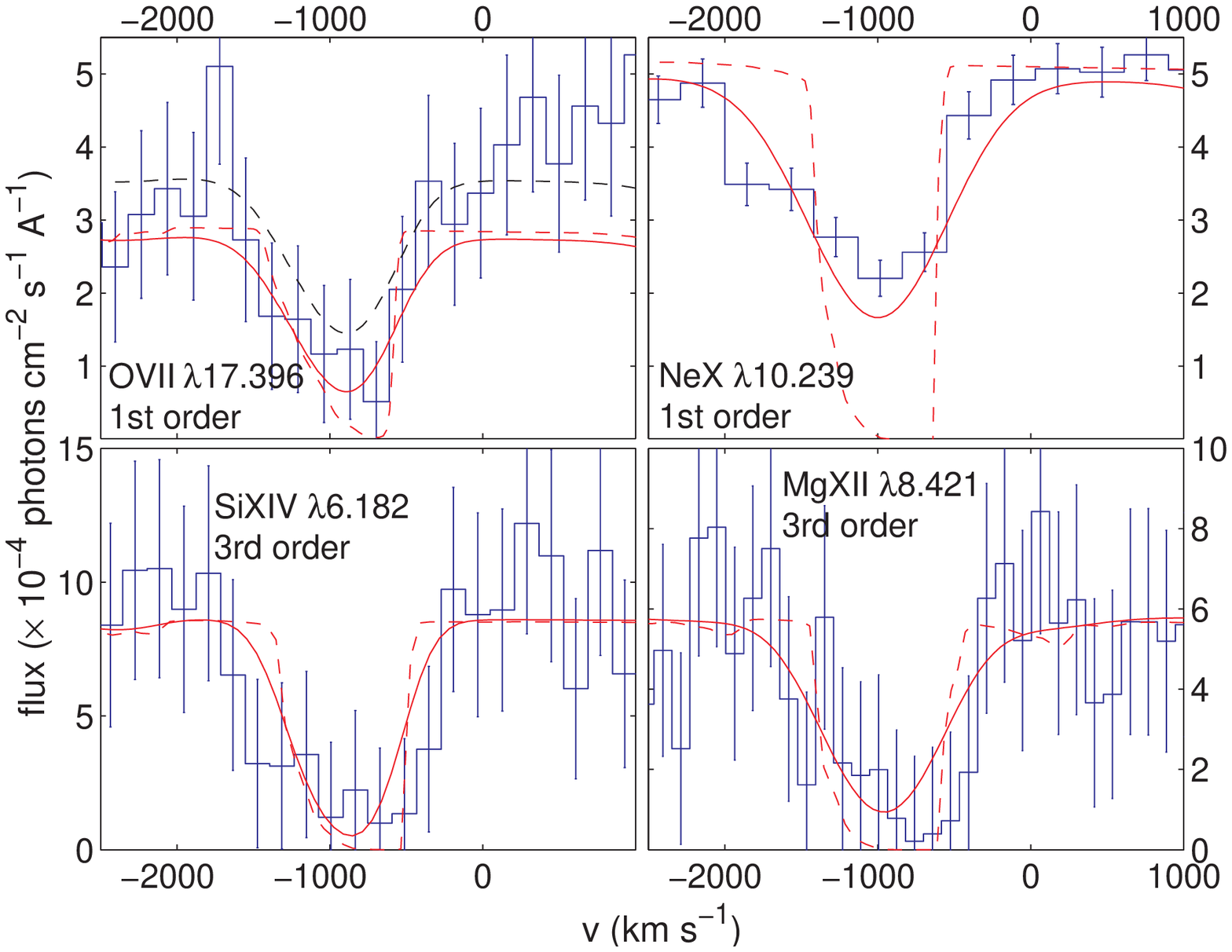}
\caption{Best-fit model to the MEG and HEG 1st and the MEG 3rd order
  absorption line profiles (model parameters are given in table
  2). The spectral model (solid line) was convolved with the instrumental
  resolution relevant to each case. Note the clear blue
  asymmetry of all predicted, non-convolved line profiles (dashed red-lines).}
\label{f15}
\end{figure}

The best-fit isothermal model is shown in figure \ref{f14} (a similar
fit is obtained for models using either of the full line-of-sight
coverage methods described in \S 4.2). Clearly, the model is
acceptable and is qualitatively similar to the best-fit model of
N03. Inspection of individual absorption line profiles (figure
\ref{f15}) shows qualitative agreement with respect to the line
centroids and widths. The derived mass loss rate  is $\dot{M}\sim
10^{-0.8\pm 0.6}C~{\rm M_\odot~yr^{-1}}$ [for $C\simeq 0.2$ (N03), $\dot{M}\sim 0.01-0.1~{\rm M_{\odot}~yr^{-1}}$], i.e., considerably smaller
than the N03  and the Behar et al. (2003) estimates. The model also
gives detailed predictions for the column densities of various
ions. These are given in table 1 for the more abundant
elements. Clearly, the stratified structure of the flow results in a
wide range of ionization levels having comparable column densities. We
discuss the implications of these results in \S 6.2.


\begin{table*}[t]
\footnotesize
\begin{center}
{\sc TABLE 1 \\ Ion Column Densities (${\rm log (cm^{-2})}$)}
\vskip 4pt
\begin{tabular}{lcccccccccccccc}
\hline
\hline
Element & I & II & III & IV & V & VI & VII & VIII & IX & X & XI & XII
& XIII & XIV  \\
\hline
H & 16.5  & & & & & & & & & & & & &  \\
He & 13.7  & 19.3 & & & & & & & & & & & &  \\
C  & 7.4 &  12.2 & 15.3 &  16.6 &   17.8 &   18.0 &  &  &  & & &
&  \\
N & 6.5 & 11.5 &   14.9 &   16.5 &   16.8 &   17.4 &   17.5 &  &
& & & & &  \\
O  & 6.8 & 11.9 &   15.5 &   17.0 &    {17.4} &   {17.5} &
{18.1} & {\bf 18.2} &    &  &  & &  \\
Ne  & 6.0 &  11.3 &   14.8 &   16.2 &   17.0 &   17.2 &   16.8 &
16.8 &   17.4 &   17.6 &   &  &  \\
Mg & 7.8 &   10.4 &   13.6 &   14.7 &   15.3 &   16.1 &   16.4 &
16.6 &   16.7 &   16.7 &   {\bf 17.1} &   {\bf 17.2} &   &  \\
Si & 6.5 &  10.2 &   12.2 &   13.4 &   14.9 &   15.8 &   {16.2} &
{16.5} &   {16.6} &   {16.7} &   {\bf 16.7} &   {\bf 16.6}
&   {\bf 17.1} & {\bf 17.2}  \\
\hline
\end{tabular}
\vskip 2pt
\parbox{4.2in}{ 
\small\baselineskip 9pt
\footnotesize
\indent
Column densities of various ions for
our best-fit model. Boldface values mark deviations by more than a
factor two from those deduced from observations and
reported by N03. 
}
\end{center}
\normalsize
\end{table*}

\begin{table}[t]
\footnotesize
\begin{center}
{\sc TABLE 2 \\ Best-fit model parameters}
\vskip 4pt
\begin{tabular}{lrr}
\hline
\hline
Parameter & Value\\
\hline
Critical distance $r_c$ [$10^{18}$\,cm] & $2.0\pm 1.5$ \\
Line-of-sight crossing distance $r_0$ [$10^{18}$\,cm] & $5\pm 2$ \\
Maximum ionization parameter at $r_0$ [${\rm log}(U_{\rm ox}^{\rm max})$] & $0.5\pm0.3$ \\
Minimum ionization parameter at $r_0$ [${\rm log}(U_{\rm ox}^{\rm min})$] & $-3.7\pm1.5$ \\
Density spectrum index  ($\beta$) & $-1.2\pm 0.5$ \\
\hline
Mass loss rate [${\rm M_{\odot}~yr^{-1}}$], & $10^{-0.8\pm
0.6}C^\star$ \\
\hline
\end{tabular}
\vskip 2pt
\parbox{4.2in}{ 
\small\baselineskip 9pt
\footnotesize
\indent
$^\star$ For $C\sim 0.2$ (N03), $\dot{M}$ is in the range $0.01-0.1~{\rm M_\odot~yr^{-1}}$
}
\end{center}
\normalsize
\end{table}

The range of  parameters characterizing acceptable  models are given
in table 2. The quoted uncertainties refer to differences between the
adiabatic and isothermal models, the mass of the central object, and the
degeneracy in the 
spectral fit. Specifically, the error in the value of each parameter
represents the maximum deviation from the best-fit value for which a
good fit can be obtained by varying all the other parameters in the
table. Thus, uncertainties in different parameters are not
independent.

\section{Discussion}


We have introduced a new approach for modeling highly ionized flows in
AGN. The emerging physical picture for the X-ray flow in NGC\,3783 is
that of a multi-phased, geometrically thick flow that is thermally
evaporated from a parsec scale region. The density range ($\sim 10^2-10^6~{\rm cm^{-3}}$ at $r_0$) occupied by the flow is much larger than obtained by 
Krongold et al. 2003 and N03 and implies a density of the warm absorbing phase which is, on average, lower than that suggested by Reynolds \& Fabian (1995) and Bottorff \& Ferland (2001). Our model has extremely ionized dynamical components that are not detected spectroscopically. The least ionized components are also absent in previous works since it is difficult to disentangle the ionization-level--column-density degeneracy.

\subsection{Is the flow turbulent?}

The above analysis suggests that the outflowing HIG may be turbulent for
several reasons: 1) The flow is multi-phased with different phases at
a rough pressure equilibrium. 2) The density spectrum (i.e., $\beta$) is
similar to that found in the ISM and in molecular clouds which are
known to be turbulent media. 3) The Reynolds number of the flow is
$Re\sim 10^7$. This is obtained by assuming non-magnetic plasma and
taking the scale of the largest eddies to be $r_0$, the velocity
$v_c$, and the density of the highest ionization phase (see e.g.,
equations 3.25 \& 3.30 in Lang 1999). Such a medium is capable of
sustaining turbulent motion.  The density  or 
range in our model spans roughly five orders of magnitude
(corresponding to $U_{\rm ox}^{\rm max}/U_{\rm ox}^{\rm min}$). This
agrees with the naive theoretical expectation that the maximum density
is that at which energy cascade is terminated and the energy is
dissipated. This occurs when the density range spans roughly  $\sim
3{\rm log}(Re)/4\simeq5$ decades (e.g., Shu 1992); i.e., consistent with the results obtained from the spectral fit. 

Our model does not explain the onset of turbulence in the flow. One possibility is that it is triggered by the putative central star cluster within the central parsec of NGC\,3783.

\subsection{The UV--X-ray connection}

Table 2 suggests that the X-ray flow in NGC\,3783 may also produce
strong low ionization UV absorption lines of \ion{H}{1}\,$\lambda
1216$, \ion{C}{4}\,$\lambda 1548$, and \ion{O}{6}\,$\lambda 1035$. Our
predictions are consistent with the lower limit on the column density
of \ion{C}{4} deduced by Crenshaw et al. (1999). The model suggests
that the global covering factor is smaller for lower ionization gas
phases (equation \ref{cg}). This, and the fact that the UV continuum
emitting source is likely to be larger than the X-ray source, imply
that the UV line profiles may be prone to partial line-of-sight
coverage effects. This may be manifested either as multi-component
absorption (i.e.,  not all flow phases are represented at every
velocity bin; see \S 4.2) or as non-black saturation (e.g., Gabel et
al. 2002). The current resolution  of the X-ray gratings and
the signal-to-noise level of the observation are not high enough to
tell which of these scenarios is relevant to NGC\,3783.

The dynamical timescale of the flow, $r_0/v_s$, is about
$10^4$\,years. Gabel et 
al. (2003) reported a considerable velocity shift ($\sim 50~{\rm
  km~s^{-1}}$) of the highest velocity component over less than a
year. Clearly, the two timescales do not agree. It is however possible
that a filament moving across our line-of-sight rather than a real
deceleration of the flow is responsible for the observed effect (see
also Gabel et al. 2003). The relevant timescale in this case is the
time it takes a filament to cross the surface of the UV emitting
region. Assuming this region to be of order $10R_s$, where $R_s$ is
the Schwartzschild radius ($\sim 10^{13}~{\rm cm}$ in the case of
NGC\,3783) and the velocity across the line-of-sight to be of the
order of the thermal speed, we get a crossing timescale  of $\sim
0.3$\,year, i.e., consistent with observations.  We conclude that
either the high velocity decelerating
UV component is not related to the X-ray flow or else the
apparent deceleration is caused by line-of-sight effects.

\subsection{Extended emission}

Some Seyfert 1 galaxies show evidence of extended, diffuse X-ray emission (e.g., NGC\,3516; George et al. 2002). This is also common in Seyfert 2 galaxies. A natural question is whether the extended HIG flow is partly responsible for such emission  through the scattering of continuum photons into our line-of-sight. The expected scattered X-ray luminosity assuming pure electron scattering is
\begin{equation}
L^x_{\rm  scat}\simeq \tau_e C L^x_{\rm tot},
\end{equation}
where $\tau_e$ is the Compton optical depth and $L^x_{\rm tot}$ the total X-ray luminosity. It is possible to relate $\tau_eC$ to the total mass loss rate of the flow, $\dot{M}$, via 
\begin{equation}
\tau_e C=\frac{\sigma_T\dot{M}}{4\pi m_H} \int_{r_e}^{\infty} {r^{-2}
  v^{-1}} {dr},
\end{equation}
where $r_e$ is the distance beyond which $L^x_{\rm scat}$ is calculated.
This gives
\begin{equation}
\frac{L^x_{\rm scat}}{L_{\rm tot}^x}\simeq6\times 10^{-6} \left (\frac{\dot{M}}{\rm M_\odot\,yr^{-1}} \right ) \left (\frac{\rm kpc}{r_e} \right ) \left ( \frac{10^3~{\rm km\,s^{-1}}}{v} \right ).
\end{equation}
The extended ($r_e\sim0.5$\,kpc) X-ray flux from NGC\,3783 is $<10^{36}~{\rm erg\,s^{-1}}$ (George I., private communication), i.e., consistent with model predictions (see table 1).

\subsection{Relation to other AGN components}

So far, we have not considered the source of the outflowing
gas. Our model suggests that, for NGC\,3783, it is located within the
central $\sim 1$\,pc. A possible mass reservoir on such scales is the
putative molecular torus (e.g., Krolik \& Begelman 1988; Krolik \&
Kriss 1995; and Netzer et al. 2002). Taking the torus parameters from
Netzer et al. (2002), we obtain that the torus mass is $\sim 10^5~{\rm
  M_\odot}$ and that HIG flows may be sustained at their current mass
loss rates for $\sim 10^6-10^8$\,years. If, as suggested by Krolik \&
Begelman (1988), the torus is also responsible for feeding the AGN,
than this timescale also corresponds to the present activity life-time.

The narrow line region (NLR) lies at the outskirts of the HIG flow and
hence is exposed to a modified  type-I AGN continuum. Similar conclusions (albeit with a somewhat more diminished UV fluxes compared to the model presented here) have been reached by Alexander et al. (1999) based on the analysis of narrow emission lines in NGC\,4151. A physical interaction between the HIG outflow and the NLR is unlikely to be energetically important given the small mass loss rate derived here (cf. Laor 1998).

\subsection{Model limitations and extensions}

Despite the success of our model in explaining the observed X-ray features there are several notable problems.

\subsubsection{flow dynamics}

In its simplest form (without turbulent line-broadening or dissipative
heating) our model under-predicted the line widths and blueshifts. One
possibility, discussed in \S 5 and \S 6.1, is that the flow is
turbulent. Alternatively, higher flow velocities and better agreement
with the data may be obtained if the flow is launched closer in to the
central source. This requires, however, that the mass of the central
object be smaller by a factor of a few compared to that reported in
Peterson et al. (2004; see also \S 6.5.2). Another possibility is that
the flow dynamics 
reflects a higher mean past luminosity compared with the one deduced
from the present observations. In
both cases, adiabatic cooling is less important (see \S 3.4) and
radiative acceleration is more effective (since $\Gamma$ is larger). Our
calculations show that in such cases the terminal velocity is 
slightly larger (by $\sim 300~{\rm km~s^{-1}}$) than that of
the isothermal model (see \S 5) and the fit is qualitatively
similar. 

\subsubsection{flow location}

The value of  $r_0$ derived here is smaller  than the lower limit
derived by N03 on the distance of
the multi-phased flow but is larger than the upper limit  for the 
highest ionization component by itself. Nevertheless, the model is consistent
with the lack of   
appreciable equivalent width (EW) variations  of the silicon and
sulfur lines since the multi-phase flow is geometrically
thick. Note also that $r_c$ and $r_0$ depend linearly on the
mass of the central object  which is somewhat uncertain. For example,
Kaspi et al. (2000) report $M_{\rm BH}\sim 10^7~M_{\odot}$, i.e., a
factor three lower than the new Peterson et al. (2004) estimate. A model
having similar spectral properties would have $r_c\sim 0.13$\,pc and
$r_0\sim 0.5$\,pc (compared with the present $r_c=2$\,pc and
$r_0=5$\,pc)  for such low $M_{\rm BH}$ (equation \ref{evap}).

Reeves et al. (2004) reported {\it XMM-Newton} Epic-pn observations
showing significant EW variations of \ion{Fe}{23}\ and \ion{Fe}{24}\
absorption lines near 2\AA. The lines are consistent with the highest
ionization N03 model yet, according to Reeves et al, they suggest the
gas lies within the central 0.02\,pc. We have checked
this suggestion within the general N03 model.  Assuming the reported line
variability is real and using the mean N03 continuum luminosity, we
find that  the  photoionization timescale of
\ion{Fe}{23} gives an upper limit of $\sim 0.2$\,pc, basically
identical to the value denoted in N03. This $r_0$  is, however, inconsistent 
with the lower-limits on the location of the two more neutral, higher density components. Thus, the Reeves et al. (2004) observation presents a real
 difficulty to our multiphase model where all components are present
 at all locations. It remains to be seen whether the high ionization
 gas component reported by Reeves et al. is part of the outflow in
 NGC\,3783.

\subsubsection{thermal instability}

Our model assumes a continuous range of densities and therefore
ionization levels and temperatures. However, some  phases of our model
lie on thermally unstable branches of the stability curve (see figure
\ref{cool}). In this case the gas may undergo non-linear compression
or expansion following some initial perturbation and end up in a
thermally stable equilibrium. This will effectively change the assumed 
$\rho-\xi$ distribution. The appropriate modeling of such conditions is beyond the scope of this paper. 

\subsubsection{spectral deviations}

Our model predicts lower flux levels then observed at soft X-ray
energies. This discrepancy is eliminated if one takes into account
partial covering (N03). Such an effect is, in fact, expected since the
low ionization gas is likely to have lower global coverage. A
continuum leakage of 20\% (80\% flow coverage) improves our fit and
correctly predicts the depth of the (saturated but non-black)
\ion{O}{8} lines. Another discrepancy is evident near $15-16$\AA\
where our model predicts too much absorption by highly ionized iron
lines. This may be related to the di-electronic recombination rates of
some of these ions (cf. Netzer 2004). Also, figure \ref{f14} does not include emission features from the flow which results in under prediction of the flux in several bands (e.g., near the $L\alpha$ line of \ion{Ne}{9} at 13.6\AA; see Kaspi et al. 2002). 

Kaspi et al. (2002) showed that some X-ray line profiles exhibit two
absorption components. Our model assumes a smooth flow and, therefore,
smooth line profiles. Multi-component absorption may be accounted for
in two ways: 1) It is possible that the high velocity component seen
in Kaspi et al. (2002) is due to a flow which is launched closer in to the
central source where the temperature of the highest ionization
component is higher  and the escape velocity larger. 2) Absorption
deficit at  some intermediate velocities (e.g., due to a ``missing''
phase at some location; see \S 6.2.2). Modeling such multi-velocity
flows is beyond the scope of this paper.

\section{Summary}

We presented a novel method for modeling the properties of highly
ionized flows in AGN. We performed state-of-the-art photoionization,
dynamical, and spectral calculations, and compared our results  to the
900\,ksec {\it Chandra/HETG} spectrum of NGC\,3783. The model is
consistent with several independent observational constraints
including the continuum and line spectrum, spectral variations (or the
lack of), and the upper-limit on  extended emission from this
source. Our major conclusions can be summarized as follows: 1) The
flow in this source is geometrically thick and is driven primarily by
thermal pressure gradients. 2) The flow is multi-phased with a
density-size spectrum which resembles that of a turbulent
medium. Adiabatic cooling and turbulent heating are likely to be
important in the energy balance. 3) The gas is located beyond the
broad line region and within the narrow line region. Its origin may be
related to the putative torus. 4) The mass loss rate is of order of
the mass accretion rate in this object and the kinetic luminosity is very small ($<0.1\%$) compared to the bolometric luminosity of the source. 5) The same flow may also
account for some of the UV lines observed in this source. 

\acknowledgments
We thank Shai Kaspi for invaluable help in dealing with the complex
{\it Chandra} spectrum of NGC\,3783. We thank N. Murray and J. Everett
for  fruitful discussions. The referee is thanked for his/her very constructive comments. While at Tel-Aviv University, D.C. was partly supported by the Dan-David Prize Foundation through the Dan-David Prize scholarship. This research has been supported by Israel Science Foundation grant no. 232/03 and by NASA through a Chandra Postdoctoral Fellowship award PF4-50033.

\appendix

\section{Alternative derivation of the flow equations}

It is possible to derive the flow equations using the volume filling
factor formalism adopted by Arav, Li, \& Begelman (1994) in their modeling of outflowing BAL regions. The Arav et al.  model assumes confined clouds and  may not be suitable for describing the highly ionized gas condensations considered here. Nevertheless, a comparison between the two approaches is useful.

Define a  volume filling factor by
\begin{equation}
\epsilon(\rho)=\epsilon_{\rm max}\left (\frac{\rho}{\rho_{\rm min}} \right )^\eta
\label{ev}
\end{equation}
We require that all flow phases fill the entire available volume, i.e., $\int d\epsilon=1$, to obtain 
\begin{equation}
\epsilon_{\rm max}=\left [ 1- \left ( \frac{\rho_{\rm max}}{\rho_{\rm min}} \right )^\eta \right ]^{-1}.
\label{e0}
\end{equation}
For a divergence free spherically expanding flow,
\begin{equation}
\dot{M}=4\pi r^2 v\int_{\rho_{\rm min}}^{\rho_{\rm min}} d\rho \epsilon(\rho)=\frac{4\pi r^2\rho_{\rm min} v}{\eta+1}\frac{1-(\rho_{\rm max}/\rho_{\rm min})^{\eta+1}}{1-(\rho_{\rm max}/\rho_{\rm min})^{\eta}}= \rm const.,
\label{mdot3}
\end{equation}
which, for a flow with a fixed density contrast, results in
\begin{equation}
\rho_{\rm min} r^2v={\rm const}.
\label{cont3}
\end{equation}
This is identical to equation \ref{cont2} and suggests that the continuity condition is not due to our choice of formalism.

Consider now a volume $V$ filled with a number of different density phases where a phase of density $\rho$ occupies $\epsilon(\rho)$ of the volume. If that phase is comprised of $n$ identical spherical clouds then the volume of each cloud is $n^{-1}\epsilon(\rho) V$. The total surface covered by such clouds is $\simeq n\times \left (n^{-1}\epsilon V )\right )^{2/3}=n^{1/3} \left ( \epsilon V \right )^{2/3}$. Thus, the global ($4\pi$) covering factor due to this phase is $C_g(\rho)\propto n^{1/3}\epsilon(\rho)^{2/3}$. 

\end{document}